\documentstyle[12pt]{article}
\begin{document}

\def\ds{\displaystyle}
\def\beq{\begin{equation}}
\def\eeq{\end{equation}}
\def\bea{\begin{eqnarray}}
\def\eea{\end{eqnarray}}
\def\beeq{\begin{eqnarray}}
\def\eeeq{\end{eqnarray}}
\def\ve{\vert}
\def\vel{\left|}
\def\brll{B\rar\rho \ell^+ \ell^-}
\def\ver{\right|}
\def\nnb{\nonumber}
\def\ga{\left(}
\def\dr{\right)}
\def\aga{\left\{}
\def\adr{\right\}}
\def\lla{\left<}
\def\rra{\right>}
\def\rar{\rightarrow}
\def\nnb{\nonumber}
\def\la{\langle}
\def\ra{\rangle}
\def\ba{\begin{array}}
\def\ea{\end{array}}
\def\tr{\mbox{Tr}}
\def\ssp{{\Sigma^{*+}}}
\def\sso{{\Sigma^{*0}}}
\def\ssm{{\Sigma^{*-}}}
\def\xis0{{\Xi^{*0}}}
\def\xism{{\Xi^{*-}}}
\def\qs{\la \bar s s \ra}
\def\qu{\la \bar u u \ra}
\def\qd{\la \bar d d \ra}
\def\qq{\la \bar q q \ra}
\def\gGgG{\la g^2 G^2 \ra}
\def\q{\gamma_5 \not\!q}
\def\x{\gamma_5 \not\!x}
\def\g5{\gamma_5}
\def\sb{S_Q^{cf}}
\def\sd{S_d^{be}}
\def\su{S_u^{ad}}
\def\ss{S_s^{??}}
\def\ll{\Lambda}
\def\lb{\Lambda_b}
\def\sbp{{S}_Q^{'cf}}
\def\sdp{{S}_d^{'be}}
\def\sup{{S}_u^{'ad}}
\def\ssp{{S}_s^{'??}}
\def\sig{\sigma_{\mu \nu} \gamma_5 p^\mu q^\nu}
\def\fo{f_0(\frac{s_0}{M^2})}
\def\ffi{f_1(\frac{s_0}{M^2})}
\def\fii{f_2(\frac{s_0}{M^2})}
\def\O{{\cal O}}
\def\sl{{\Sigma^0 \Lambda}}
\def\es{\!\!\! &=& \!\!\!}
\def\ar{&+& \!\!\!}
\def\ek{&-& \!\!\!}
\def\cp{&\times& \!\!\!}
\def\se{\!\!\! &\simeq& \!\!\!}
\def\hml{\hat{m}_{\ell}}
\def\rr{\hat{r}_{\Lambda}}
\def\ss{\hat{s}}


\renewcommand{\textfraction}{0.2}    
\renewcommand{\topfraction}{0.8}

\renewcommand{\bottomfraction}{0.4}
\renewcommand{\floatpagefraction}{0.8}
\newcommand\mysection{\setcounter{equation}{0}\section}

\def\baeq{\begin{appeq}}     \def\eaeq{\end{appeq}}
\def\baeeq{\begin{appeeq}}   \def\eaeeq{\end{appeeq}}
\newenvironment{appeq}{\beq}{\eeq}
\newenvironment{appeeq}{\beeq}{\eeeq}
\def\bAPP#1#2{
 \markright{APPENDIX #1}
 \addcontentsline{toc}{section}{Appendix #1: #2}
 \medskip
 \medskip
 \begin{center}      {\bf\LARGE Appendix #1 :}{\quad\Large\bf #2}
\end{center}
 \renewcommand{\thesection}{#1.\arabic{section}}
\setcounter{equation}{0}
        \renewcommand{\thehran}{#1.\arabic{hran}}
\renewenvironment{appeq}
  {  \renewcommand{\theequation}{#1.\arabic{equation}}
     \beq
  }{\eeq}
\renewenvironment{appeeq}
  {  \renewcommand{\theequation}{#1.\arabic{equation}}
     \beeq
  }{\eeeq}
\nopagebreak \noindent}

\def\eAPP{\renewcommand{\thehran}{\thesection.\arabic{hran}}}

\renewcommand{\theequation}{\arabic{equation}}
\newcounter{hran}
\renewcommand{\thehran}{\thesection.\arabic{hran}}

\def\bmini{\setcounter{hran}{\value{equation}}
\refstepcounter{hran}\setcounter{equation}{0}
\renewcommand{\theequation}{\thehran\alph{equation}}\begin{eqnarray}}
\def\bminiG#1{\setcounter{hran}{\value{equation}}
\refstepcounter{hran}\setcounter{equation}{-1}
\renewcommand{\theequation}{\thehran\alph{equation}}
\refstepcounter{equation}\label{#1}\begin{eqnarray}}


\newskip\humongous \humongous=0pt plus 1000pt minus 1000pt
\def\caja{\mathsurround=0pt}


\title{
 {\small \begin{flushright}
\today
\end{flushright}}
       {\Large
                 {\bf
 Exclusive $B \rar \rho \ell^+ \ell^-$ Decay in the Standard Model with Fourth--Generation
                 }
         }
      }

\author{\vspace{1cm}\\
{\small  K. Zeynali$^1$\thanks {e-mail: k.zeinali@arums.ac.ir}\,\,,
V. Bashiry$^2$\thanks {e-mail: bashiry@ipm.ir}\,\,,
} \\
{\small $^1$Faculty of Medicine,
Ardabil University of Medical Sciences (ArUMS) ,}\\{ \small  Daneshgah St., Ardabil, Iran }\\
{\small $^2$  Cyprus International University, Via Mersin 10,
Turkey }\\}
\date{}
\begin{titlepage}
\maketitle
\thispagestyle{empty}

\begin{abstract}
We investigate the influence of the fourth generation of quarks on
the branching ratio, the CP-asymmetry and the polarization
asymmetries in $B \rar \rho \ell^+ \ell^-$ decay. Taking
$|V_{t'd}V_{t'b}|\sim 0.001$ with phase about $10^\circ$, which is
consistent with the $sin2\phi_1$ of the CKM and the $B_d$ mixing
parameter $\Delta m_{B_d}$, we obtain that for both ($\mu, \,
\tau$) channels the branching ratio is increased and the magnitude
of CP-asymmetry and polarization asymmetries decreased by the mass
and mixing parameters of the 4th generation of quarks . These
results can serve as a good tool to search for new physics
effects, precisely, to search for the fourth generation of
quarks($t',\, b')$ via its indirect manifestations in loop
diagrams.
\end{abstract}

~~~PACS numbers: 12.15.Ji, 13.25.Hw
\end{titlepage}

\section{Introduction}
Flavor changing natural current (FCNC) and lepton flavor violation
(LFV) are forefront of our study both for precision test of the
Standard Model (SM) and for new physics effects. FCNC, forbidden
in the tree level, is induced by quantum loop level. The new
physics(NP) can either contribute to the effective Hamiltonian by
the new operators which are absent in the SM or alter the Wilson
coefficients of the Hamiltonian. A consequential extension of the
SM with new generation of fermions belongs to the classes of the
new physics where the Wilson coefficients  change comparing to the
corresponding three--generation Standard Model(SM3).

The existence of the  4th generation of fermions, if their mass is
less than the half of the mass of the Z boson, is excluded by the
LEP II experiment\cite{LEP}. In this sense, the status of the
fourth generation is more subtle \cite{Frampton} from the
experimental point of view. However, a consequential extension of
the SM3 can address some of the puzzles and fundamental questions
from the theoretical point of view. In this respect, the
consequential 4th generation of quarks and leptons are interesting
in different ways i.e., \cite{Bashiry2007}--\cite{Arhrib:2002md}.
The 4th generation of quarks can include new weak phases and
mixings in the Cabibbo-Kobayashi-Maskawa matrix(CKM). Thus, the
four-generation Standard Model(SM4) can demonstrate a better
solution to the baryogenesis than the SM3.

 Two type of studies can be conducted to
discover the 4th generation of fermions. The first type  is the
direct search of the 4th generation of quarks and leptons which
can be accessed by increasing the center of mass energy of
colliders with high luminosity. Here the cross section of
production will increase and such fermions can be created as real
states; i.e., the 4th generation of quarks can be created by
gluon--gluon fusion at LHC\cite{Arik:1996qd}. The second type is
the indirect search dealing with the effects of the 4th generation
of fermion in the FCNC
decays~\cite{Bashiry2007}--\cite{Arhrib:2002md} and
LFV~\cite{Wu-Jun}. In these classes of studies, one studies the
contribution of the 4th generation of fermions at the quantum loop
level; ref.~\cite{Wu-Jun} studied the effects of the 4th
generation of heavy neutrino (heavier than the half of the Z boson
mass) in the $\mu \to e \gamma $ decay and anomalous magnetic
moment of the $\mu$. The result was an upper limit  for the mass
of $\nu_4$ which is up to $\sim 100$GeV . Considering these
constrains, one can study the branching ratio of the $\mu^- \to
2e^-e^+$ decay.

 The $b \rar s(d)$ transition is forefront for searching the 4th generation of quarks. This
transition is forbidden at tree level in the Standard Model. A
consequential extension of three-generation Standard Model to the
four-generation Standard Model (SM4) maintains the same property
at tree level, but at the quantum loop level the 4th generation of
heavy quark $(t')$ can contribute to the quantum loop. This
contribution can affect physical observables, i.e. branching
ratio, CP asymmetry, polarization asymmetries and
forward--backward (FB) asymmetries. The study of these physical
observables is a good tools to look for the 4th generation of up
type quarks\cite{Bashiry2007}--\cite{Arhrib:2002md}.

There are  some constraints on a fourth
family\cite{Sultansoy:2000dm}. From the strong constraint on the
number of light neutrinos we know that the fourth family of
neutrino is heavy. The S and $\rho$ parameter are sensitive to a
fourth family, but the experimental limits on these parameters
have been evolved through years in such a way that the constraint
on a fourth family has lessened. In addition, the masses of the
fourth family of leptons may produce negative S and T. As
discussed in \cite{4thgood} and the reference therein, the
constraints from S and T do not prohibit the fourth family, but
instead serve only to constrain the mass spectrum of the fourth
family of quarks and leptons.

 FCNC and CP--violation (CPV)
are indeed the most sensitive probes of NP contributions to
penguin operators. Rare decays, induced by flavor changing neutral
current (FCNC) $b \rar s(d)$ transitions, are at the forefront of
our quest to understand flavor and the origins of CPV, offering
one of the best probes for new physics beyond the Standard Model
(SM) \cite{R23,Willey,deshpande}. In addition, there are some
important QCD corrections, which have recently been calculated in
the NNLO\cite{NNLL}. Moreover, $b \rar s(d) \ell^+ \ell^-$ decay
is also very sensitive to the new physics beyond SM. New physics
effects manifest themselves in rare decays in two different ways,
either through new combinations to the new Wilson coefficients or
through the new operator structure in the effective Hamiltonian,
which is absent in the SM. A crucial problem in the new physics
search within flavour physics is the optimal separation of new
physics effects from uncertainties. It is well known that
inclusive decay modes are dominated by partonic contributions;
non--perturbative corrections are in general smaller~\cite{Hurth}.
Also, ratios of exclusive decay modes such as asymmetries for
$B\rar K(~K^\ast,~\rho,~\gamma)~ \ell^+ \ell^-$ decay
\cite{R4621}--\cite{bashirychin} are well studied for new physics
search. Here large parts of the hadronic uncertainties partially
cancel out.

In this paper we investigate the possibility of searching for new
physics in the  $\brll$ decay using the SM with the fourth
generation of quarks ($b',\, t'$). The fourth quark ($t'$), like
$u,c,t$ quarks, contributes in the $b \rar s(d) $ transition at
loop level. Clearly, it would change the branching ratio,
CP-asymmetry and polarization asymmetries. Note that fourth
generation effects on the branching ratio have been widely studied
in baryonic and semileptonic $b\rar s$ transition
\cite{Huo:2003cj,Aliev:2003gp},
\cite{Hou:2006jy}--\cite{London:1989vf}. But few studies related
to the $b\rar d $ transitions~\cite{Bashiry2007} exist.

The sensitivity of the branching ratio and CP asymmetry  to the
existence of the fourth generation of quarks in the $B \rar \pi
\ell^+ \ell^-$ decay is investigated in \cite{Bashiry2007} and it
was observed that branching ratio, CP asymmetry and lepton
polarization asymmetries are very sensitive to the fourth
generation parameters ($m_{t'}$, $V_{t'b}V^*_{t'd}$ ). In this
regard it is interesting to ask whether the branching ratio,
CP-asymmetry and lepton polarization asymmetries in $B \rar \rho
\ell^+ \ell^-$ decay are sensitive to the fourth generation
parameters in the same way.
 In the work  presented here we try to answer  these questions.

The paper is organized as follows: In section 2, using the
effective Hamiltonian, the general expressions for the matrix
element and CP asymmetry of $ B \rar \rho \ell^+ \ell^-$  decay is
derived. Section 3 is devoted to calculations of lepton
polarization. In section 4, we investigate the sensitivity of
these functions to the fourth generation parameters ($m_{t'}$,
$V_{t'b}V^*_{t'd}$ ).

\section{Matrix Element Differential Decay Rate and CP Asymmetry}

The QCD corrected effective Lagrangian for the decays
$b\rightarrow d\ell^{+}\ell^{-}$ can be achieved by integrating
out the heavy quarks and the heavy electroweak bosons in the SM4
as follows:

\begin{eqnarray}\label{e1}
 M =\frac{G_{F}\alpha_{em}
\lambda_t}{\sqrt{2}\pi}[&C^{\rm
tot}_{9}(\overline{d}\gamma_{\mu}P_{L}b)\overline{\ell}\gamma_{\mu}\ell
+C_{10}^{\rm
tot}(\overline{d}\gamma_{\mu}P_{L}b)\overline{\ell}\gamma_{\mu}\gamma^{5}\ell
\nonumber\\&-2\,C_{7}^{\rm
tot}\overline{d}i\sigma_{\mu\nu}\frac{q^{\nu}}{q^{2}}
(m_{b}P_{R}+m_{d}P_{L})b\overline{\ell}\gamma_{\mu}\ell\,]
 , \,\,\, \label{amplitude}
\end{eqnarray}
In this formula unitarity of the CKM matrix has been used. Here
the $\lambda_t=V_{tb}^\ast V_{td}$ is factored out and
 $q$ denotes the four momentum of the lepton pair. The  Wilson coefficients $C_i^{\rm tot}$'s are as follows:
 \bea\label{ctot}\lambda_t C_i^{tot}=\lambda_t C^{SM}_i+\lambda_{t'} C^{new}_i~,\eea hereby, $\lambda_f=V_{fb}^\ast V_{fd}$ and the last
term in these expression describes the contributions of the
$t^\prime$ quark to the Wilson coefficients. The explicit forms of
the $C_i^{\rm new}$ can be obtained from the corresponding
expression of the Wilson coefficients in the SM by substituting
$m_t \rar m_{t^\prime}$.

 A general $4\times4$ CKM matrix can be written as
follows:
\begin{equation}\label{standard}
\hat V^4_{\rm CKM}=\left(\begin{array}{cccc}
V_{ud}&V_{us}&V_{ub}&V_{ub'}\\ V_{cd}&V_{cs}& V_{cb}&V_{cb'}\\
V_{td}&V_{ts}&V_{tb}&V_{tb'}\\V_{t'd}&V_{t's}&V_{t'b}&V_{t'b'}
\end{array}\right)
\end{equation}
Using the  Wolfenstein parametrization,  the values of $3\times3$
CKM matrix elements, keeping ${\cal O}(\lambda^{5})$, is obtained
in $\cite{BurasB}$. On the other hand, one can estimate the
elements appearing in the fourth column and row of $4\times4$ CKM
matrix by studying  the experimental results of the $B_{s,d}$
mixing\cite{Hou:2006jy}  and $b\to s(d)$
transitions\cite{Hou:2005hd, Zolfagharpour:2007eh}. The former
sharply constrains the phases of elements and the latter generally
constrains the magnitudes. If we summarize all these experimental
constrains with the unitarity condition  of $4\times4$ CKM matrix,
then the following values for the elements of $\hat{V}^4_{\rm
CKM}$ can be obtained: \bea\label{Hou:2005hd} \hat V^4_{\rm CKM}
\approx \left(\begin{array}{cccc}
0.9745&0.224&0.0038e^{-i 60^\circ}&0.0281e^{i 61^\circ}\\ -0.224&0.9667& 0.0415&0.1164e^{i 66^\circ}\\
0.0073e^{-i25^\circ}&0.0555e^{-i25^\circ}&0.9746&0.2168e^{-i1^\circ}\\-0.0044e^{-i10^\circ}&-0.1136e^{-i70^\circ}&-0.2200&0.9688
\end{array}\right),
\eea The unitarity of the $4\times4$ CKM matrix leads to
\bea\label{unitarity}\lambda_u+\lambda_c+\lambda_t+\lambda_{t'}=0.\eea
 Then
$\lambda_t=-\lambda_u-\lambda_c-\lambda_{t'}$.
 Now we can
re-write Eq.~\ref{ctot} as:  \bea \lambda_t
C^{SM}_{7,10}+\lambda_{t'} C^{new}_{7,10}=(-\lambda_c-\lambda_u)
C^{SM}_{7,10}+\lambda_{t'} (C^{new}_{7,10}-C^{SM}_{7,10} )\eea It
is clear that for the $m_{t'}\rar m_t$ or $\lambda_{t'}\rar 0$ the
$\lambda_{t'} (C^{new}_{7,10}-C^{SM}_{7,10} )$ term vanishes and
the SM3 results are recovered.
 If we parameterize $\lambda_{t'}$ as: \bea
{\label{parameter}} \lambda_{t'}=V_{t^\prime b}^\ast V_{t^\prime
d}=r_{db}e^{i\phi_{db}}\eea it is obvious from Eq.
(\ref{Hou:2005hd}) that $\phi_{db}\sim 10^0$ and $r_{db}\sim{\cal
O}(10^{-3})$.

 Neglecting the terms of
$O(m_q^2/m_W^2)$, $q = u, d, c$, the analytic expressions for all
Wilson coefficients, except $C_{9}^{eff}$,  can be found in
\cite{burev}. Note that just $C_9^{\rm tot}$ has weak and strong
phases, i.e.:
\begin{equation}\label{c9new}
C^{\rm tot}_{9}=\xi_{1}+\lambda_{tu}\xi_{2}+\lambda_{tt'}C^{\rm
new}_9
\end{equation}
where the CP violating parameter
$\lambda_{tt'}=\frac{\lambda_{t'}}{\lambda_{t}}$ and $
  \lambda_{tu}=\frac{\lambda_{u}}{\lambda_{t}}$.

 The explicit expressions of functions
$\xi_1$ and $\xi_2$ in $\mu=m_b$ scale  are
respectively~\cite{burev}--\cite{misiakE}:
\begin{eqnarray}\label{xi1}
\xi_1 & = & C_9(x_i,m_b) \, +\, 0.138
\omega(\hat{s})\,+\,g(\hat{m}_{c},\hat{s}) (3 C_1 + C_2 + 3 C_3 +
C_4 + 3 C_5 + C_6)\nonumber\\&-& \frac{1}{2}g(\hat{m}_{d},\hat{s})
(C_3 + C_4) - \frac{1}{2}
   g(\hat{m}_{b},\hat{s})(4 C_3 + 4 C_4 + 3C_5 + C_6) \\
  & + &\frac{2}{9} (3 C_3 + C_4 + 3C_5 + C_6)\nonumber \\
   \xi_2 & = & [ g(\hat{m}_{c},\hat{s})- g(\hat{m}_{u},\hat{s})](3
C_1 + C_2)\label{xi2}
\end{eqnarray}
where $\hat{m}_q=\,m_q/m_b$, $\hat{s}=\,\frac{q^2}{m_b^2} $, and
$x_i=\frac{m_i^2}{m_W^2}$ ,here, $i=t(t')$ for $C_9^{eff}(C_9^{\rm
new})$.

The function $g(\hat{m}_{q}, \hat{s})$, which includes strong
phase, represents the one loop corrections to the four-quark
operators $O_1-O_6$ \cite{misiakE} and is defined as:
\begin{eqnarray}
g(\hat{m}_{q}, \hat{s}) &=& -\frac{8}{9} \ln(\hat{m}_{q}) +
\frac{8}{27} + \frac{4}{9}\ y_{q}
          - \frac{2}{9} (2 + y_{q}) \sqrt{|1-y_{q}|}\
\bigg\{\Theta(1-y_{q}) \times
\nonumber\\
&&\left[ \ln\left(\frac{1+\sqrt{1-y_{q}}}{1-\sqrt{1-y_{q}}}\right)
- i \pi \right]
 + \Theta(y_{q} - 1)\ 2 \arctan{\frac{1}{\sqrt{y_{q}-1}}}
\bigg\}~, \label{gmq}
\end{eqnarray}
Although long-distance effects of $c\bar{c}$ bound states could
contribute to $C^{\rm eff}_9$, for simplicity, they are not included
in the present study. On the other hand, the bound states could be
excluded experimentally by cutting the phase space at the resonant
regions. In the case of  the $J/\psi$ family, this is usually
accomplished by introducing a Breit-Wigner distribution for the
resonances through the replacement \cite{longdist}
 \begin{eqnarray}\label{long-dist}
  g(\hat{m}_{c}, \hat{s}) \longrightarrow
g(\hat{m}_{c}, \hat{s})-\frac{3\pi}{\alpha^2}\sum\limits_{V=
J/\psi, \psi', \dots} \frac{\hat{m}_V Br(V\to
l^+l^-)\hat{\Gamma}_{\mathrm {total}}^V} {\hat{s}-\hat{m}^2_V + i
\hat{m}_V \hat{\Gamma}_{\mathrm {total}}^V}.
\end{eqnarray}

One has to sandwich the inclusive effective Hamiltonian between
initial hadron state $B(p_B)$ and final hadron state
$\rho(p_{\rho})$ to obtain the matrix element for the exclusive
decay $B (p_{B}) \to \rho(p_{\rho})~ \ell^+ (p_+)\ell^-(p_-)$. It
follows from Eq.~(\ref{e1}) that in order to calculate the decay
width and other physical observables of the exclusive $B \rar \rho
\ell^+ \ell^-$ decay, we need the following  matrix elements,
defined in terms of formfactors~\cite{Colangelo}:
\bea\label{form1}
\langle\rho(p_\rho,\varepsilon)|\bar{d}\gamma_\mu(1-\gamma^5)b|B(p_B)\rangle&=&-\epsilon_{\mu\nu\lambda\sigma}
\varepsilon^{\nu*}p^\lambda_\rho
p^\sigma_B\frac{2V(q^2)}{m_B+m_\rho}-i\varepsilon_\mu^*(m_B+m_\rho)A_1(q^2)\nnb
\\  &+&
i(p_B+p_\rho)(\varepsilon^*q)\frac{A_2(q^2)}{m_B+m_\rho}\nnb\\&+&iq_\mu(\varepsilon^*q)\frac{2m_\rho}{q^2}[A_3(q^2)
-A_0(q^2)],\\
\label{form2}\langle\rho(p_\rho,\varepsilon)|\bar{d}i\sigma_{\mu\nu}q^\nu(1\pm\gamma^5)b|B(p_B)\rangle&=&
4\epsilon_{\mu\nu\lambda\sigma}\varepsilon^{\nu*}p^\lambda_\rho
q^\sigma T_1(q^2)\pm2i[\varepsilon^*_\mu(m_B^2-m^2_\rho)\nnb
\\  &-&
(p_B+p_\rho)_\mu(\varepsilon^*q)]T_2(q^2)\nnb\\&\pm&2i(\varepsilon^*q)
\Bigg(q_\mu-(p_B+p_\rho)_\mu\frac{q^2}{m_B^2-m_\rho^2}\Bigg)T_3(q^2),
 \\
\langle\rho(p_\rho,\varepsilon)|\bar{d}(1+\gamma^5)b|B(p_B)\rangle&=&-\frac{2im_\rho}{m_b}(\varepsilon^*q)A_0(q^2),\label{form3}\eea
 where $p_\rho$ and
 $\varepsilon$
  denote the four momentum and
 polarization vector of the $\rho$
  meson, respectively.

 From Eqs.~(\ref{form1},\ref{form2},\ref{form3}) we get the following
 expression for the matrix element of the $B \rar \rho
 \ell^+\ell^-$decay:
 \bea\label{14}
  M^{B\rightarrow\rho}&=&\frac{G_{F}\alpha_{em}
\lambda_t}{\sqrt{2}\pi}\Bigg[i\epsilon_{\mu\nu\alpha\beta}\epsilon^{\nu*}p_B^\beta
  q^\beta A+\epsilon_\mu^*B+(\epsilon^*.q)(p_B)C\Bigg](\bar{\ell}\gamma^\mu\ell)
\\ \nnb &+&
\Bigg[i\epsilon_{\mu\nu\alpha\beta}\epsilon^{\nu*}p_B^\beta
q^\beta
D+\epsilon_\mu^*E+(\epsilon^*.q)(p_B)F\Bigg](\bar{\ell}\gamma^\mu\ell)+G(\epsilon^*.q)(\bar{\ell}\gamma_5\ell)
 \eea
 where
 \bea\label{15}
A&=&\frac{4(m_b+m_d)T_1(q^2)}{m_B^2s}C_7^{tot}+\frac{V(q^2)}{m_B+m_\rho}C_9^{tot},\nnb
\\
B&=&-\frac{2(m_b-m_d)(1-r_\rho)T_2(q^2)}{s}C_7^{tot}-\frac{(m_B+m_\rho)A_1(q^2)}{2}C_9^{tot},\nnb
\\
C&=&\frac{4(m_b-m_d)}{m_B^2s}\bigg(T_2(q^2)+\frac{s}{1-r_\rho}T_3(q^2)\Bigg)C_7^{tot}+\frac{A_2(q^2)}{m_B+m_\rho}C_9^{tot},\nnb
\\
D&=&\frac{V(q^2)}{m_B+m_\rho}C_{10}^{tot},\nnb
\\
E&=&-\frac{(m_B+m_\rho)A_1(q^2)}{2}C_{10}^{tot},\nnb
\\
F&=&\frac{A_2(q^2)}{m_B+m_\rho}C_{10}^{tot},\nnb
\\
G&=&\Bigg(-\frac{m_\ell }{m_B+m_\rho}A_2(q^2)+\frac{2m_\rho
m_\ell}{m_B^2s}(A_3(q^2)-A_0(q^2))\Bigg)C_{10}^{tot}.
 \eea
From this expression of the matrix element, for the differential
decay width we get the following result: \bea\label{16}
\Bigg(\frac{d\Gamma^\rho}{ds}\Bigg)_0&=&\frac{G_F^2\alpha^2}{3\times2^{10}\pi^5}|\lambda_t|^2m_B^5v\sqrt{\lambda_\rho}\Delta_\rho,
\eea
 \bea\label{17}
 \Delta_\rho&=&(1+\frac{2t^2}{s})\lambda_\rho\Bigg[4m_B^2s|A|^2+\frac{2}{m_B^2r_\rho}(1+12\frac{sr_\rho}{\lambda_\rho})|B|^2\nnb
\\  &+&
\frac{m_B^2}{2r_\rho}\lambda_\rho|C|^2+\frac{2}{r_\rho}(1-r_\rho+s)~{\rm Re}(B^*C)\Bigg]+4m_B^2\lambda_\rho(s-4t^2)|D|^2\nnb
\\  &+&
\frac{4(2t^2+s)-4(2t^2+s)(r_\rho+s)+4t^2(r_\rho^2-26r_\rho+s^2)+2s(r_\rho^2+10sr_\rho+s^2)}{m_B^2sr_\rho}|E|^2\nnb
\\  &+&
\frac{m_B^2}{2sr_\rho}\lambda_\rho\Bigg[(2t^2+s)(\lambda_\rho+2s+2r_\rho)-2\{2t^2(r_\rho+5s)+s(r_\rho+s)\}\Bigg]|F|^2\nnb
\\  &+&
3\frac{s}{r_\rho}\lambda_\rho|G|^2+\frac{2\lambda_\rho}{sr_\rho}\Bigg[-2t^2(r_\rho-5s)+(2t^2+s)-s(r_\rho+s)\Bigg]~{\rm Re}(E^*F)\nnb
\\  &+&
\frac{12t}{m_Br_\rho}\lambda_\rho ~{\rm
Re}(G^*E)+\frac{2m_Bt}{r_\rho}\lambda_\rho(1-r_\rho+s)~{\rm
Re}(G^*F).
 \eea
 with $r_\rho=m_\rho^2/m_B^2, \lambda_\rho=r_\rho^2+(s-1)^2-2r_\rho(s+1),
 v=\sqrt{1-\frac{4t^2}{s}}$ and
$t=m_\ell/m_B.$

Another physical observable is the CP--violating  asymmetry which
can be defined for both polarized and unpolarized leptons. We aim
to obtain the normalized CP--violating asymmetry for the
unpolarized leptons. The standard definition is given as: \bea
\label{acp} A^\rho_{CP}(\hat{s}) = \frac{\ds{\ga
\frac{d\Gamma^\rho}{d\hat{s}}\dr_0}- \ds{\ga
\frac{d\bar{\Gamma}^\rho}{d\hat{s}}\dr_0} } {\ds{\ga
\frac{d\Gamma^\rho}{d\hat{s}}\dr_0}+ \ds{\ga
\frac{d\bar{\Gamma^\rho}}{d\hat{s}}\dr_0} } = \frac{\Delta_\rho
-\bar{\Delta}_\rho}{\Delta_\rho +\bar{\Delta}_\rho}~, \eea where
\bea \frac{d\Gamma^\rho}{d\hat{s}} = \frac{d\Gamma^\rho(b\rar
d\ell^+\ell^-)}{d\hat{s}},~\mbox{\rm and},~
\frac{d\bar{\Gamma}^\rho}{d\hat{s}} =
\frac{d\bar{\Gamma}^\rho(b\rar d\ell^+\ell^-)}{d\hat{s}}~,\nnb
\eea and $(d\bar{\Gamma}^\rho/d\hat{s})_0$ can be obtained from
$(d\Gamma^\rho/d\hat{s})_0$ by making the replacement \bea
\label{e5714} C_9^{tot} =
\xi_1+\lambda_{tu}\xi_2+\lambda_{tt^\prime}C_9^{\rm new} \rar
\bar{C}_9^{tot}=
\xi_1+\lambda_{tu}^\ast\xi_2+\lambda_{tt^\prime}^*C_9^{\rm new}~.
\eea Using this definition and the expression for
$\Delta^\rho(\hat{s})$, the CP violating asymmetry contributed
from SM3 and new contributions from SM4 are: \bea
A^\rho_{CP}(\hat{s})=\frac{-\Sigma^{SM}-\Sigma^{\rm new}}
{\Delta_1^{\rho}+\Sigma^{SM}+\Sigma^{\rm new}}\eea where
\bea\label{sigma-Sm}
 \Sigma^{SM}(s)&=&4~{\rm Im}(\lambda_{tu})\Bigg(B_4~{\rm Im}(\xi_1^*\xi_2)+
\frac{B_2+B_3}{2}~{\rm Im}(C_7^{\rm eff*}\xi_2) \Bigg),
 \eea
 \bea\label{sigma-new}
\Sigma^{\rm new}(s)&=&4~{\rm Im}(\lambda_{tt^\prime})\Bigg(B_1~{\rm Im}(C_7^{\rm new}C_7^{\rm eff*})+B_2~{\rm Im}(C_9^{\rm new}C_7^{\rm eff*})\nnb
\\  &+&
B_3~{\rm Im}(C_7^{\rm new}\xi_1^*) +B_4~{\rm Im}(C_9^{\rm new}\xi_1^*)\Bigg)\nnb
\\  &+&
4~{\rm Im}(\lambda_{tt^\prime}\lambda_{tu})\Bigg(\frac{B_2+B_3}{2}~{\rm Im}(C_7^{\rm eff*}\xi_2)+B_4~{\rm Im}(\xi_1^*\xi_2)\Bigg)\nnb
\\  &+&
4~{\rm Im}(\lambda_{tt^\prime}^*\lambda_{tu})\Bigg(\frac{B_2+B_3}{2}~{\rm Im}(C_7^{\rm new*}\xi_2)+B_4~{\rm Im}(C_9^{\rm new*}\xi_2)\Bigg)\nnb
\\  &+&
4~{\rm Im}(\lambda_{tu})|\lambda_{tt^\prime}|^2\Bigg(\frac{B_2+B_3}{2}~{\rm Im}(C_7^{\rm new*}\xi_2)+B_4~{\rm Im}(C_9^{\rm new*}\xi_2)\Bigg),
 \eea with
 \bea\label{B1}
B_1&=&\frac{8m_b^2}{m_B^2sr_\rho\lambda_\rho}\Bigg(2r_\rho
s(\lambda_\rho-12r_\rho+4r_\rho^2+6)+\lambda_\rho(2r_\rho-1-r_\rho^2-2s-\lambda_\rho)\Bigg)T_2^2\nnb
\\  &+&
\frac{16m_b^2}{m_B^2(1-r_\rho)r_\rho
s}\Bigg(r_\rho(s-r_\rho+2)+(\lambda_\rho-s-1)\Bigg)T_2T_3\nnb
\\  &+&
\frac{64m_b^2}{m_B^2s}T_1^2+\frac{8m_b^2\lambda_\rho}{m_B^2(1-r_\rho)^2r_\rho}T_3^2
 \eea

 \bea\label{B2}
B_2&=&\frac{2m_b}{(m_B+m_\rho)r_\rho
s}\Bigg(2r-\rho(s+r_\rho+2)+(2s+\lambda_\rho+2)\Bigg)A_2T_2\nnb
\\  &+&
\frac{2m_b}{m_B^2sr_\rho\lambda_\rho}\Bigg(\lambda_\rho(m_B+m_\rho)(1-r_\rho)+
12sr_\rho(m_\rho-r_\rho(m_N+m_\rho)+\lambda_\rho)\Bigg)A_1T_2\nnb
\\  &+&
\frac{16m_b}{(m_B+m_\rho)}T_1V+\frac{2m_b\lambda_\rho}{(m_B+m_\rho)(1-r_\rho)r_\rho}A_2T_3
\\ \label{B3}
B_3&=&B_2+\frac{4m_b(1-r_\rho+s)}{m_B^2(m_B+m_\rho)(1-r_\rho)sr_\rho}\Bigg(m_B^2(1-r_\rho)^2A_2T_2\nnb
\\  &-&
((m_B+m_\rho)^2(1-r_\rho)T_2+sT_3)A_1 \Bigg)
 \\ \label{B4}
B_4&=&\frac{1}{2m_B^2(m_B+m_\rho)^2r_\rho\lambda_\rho}\Bigg(2m_B^2(m_B+m_\rho)^2(r_\rho-s-1)\lambda_\rho
A_1A_2\nnb
\\  &+&
(m_B+m_\rho)^4(12sr_\rho+\lambda_\rho)A_1^2+m_B^4\lambda_\rho(8sr_\rho
V^2+\lambda_\rho A_2^2)\Bigg)
 \eea
 and
 \bea\label{delta1}
\Delta_1^\rho&=&\frac{2s\Delta_\rho}{(s+2t^2)\lambda_\rho}.
 \eea
 From this expression it is easy to see that in the
$\lambda_{t'}\rightarrow 0$ the SM3 result can be obtained.
Secondly, when $m_{t'}\rightarrow m_t$ the result of the SM4
coincides with the SM3, as it has to be, even if it is not obvious
visible from the expressions (see figures).
\section{Lepton polarization}
 In order to calculate the
polarization asymmetries of the lepton defined in the effective
four fermion interaction of Eq. (\ref{14}), we must first define
the orthogonal vectors (components of $S$) in the rest frame of
$\ell^-$. Note that we should use the subscripts $L$, $N$ and $T$
to correspond to the lepton being polarized along the
longitudinal, normal and transverse directions, respectively.
\begin{eqnarray}
S^\mu_L & \equiv & (0, \mathbf{e}_{L}) ~=~ \left(0,
\frac{\mathbf{p}_-}{|\mathbf{p}_-|}
\right) , \nonumber \\
S^\mu_N & \equiv & (0, \mathbf{e}_{N}) ~=~ \left(0,
\frac{\mathbf{p_{\rho}} \times \mathbf{p}_-}{|\mathbf{p_{\rho}}
\times
\mathbf{p}_- |}\right) , \nonumber \\
S^\mu_T & \equiv & (0, \mathbf{e}_{T}) ~=~ \left(0, \mathbf{e}_{N}
\times \mathbf{e}_{L}\right) , \label{sec3:eq:1} \eea

where $\mathbf{p}_-$ and $\mathbf{p_{\rho}}$ are the three momenta
of the $\ell^-$ and $\rho$ particles, respectively. The
longitudinal unit vector is boosted to the CM frame of $\ell^-
\ell^+$ by Lorenz transformation:
\begin{eqnarray}
S^\mu_L & = & \left( \frac{|\mathbf{p}_-|}{m_\ell}, \frac{E_{\ell}
\mathbf{p}_-}{m_\ell |\mathbf{p}_-|} \right) , \nonumber
\end{eqnarray}
here, the other two vectors remain unchanged. The polarization
asymmetries can now be calculated using the spin projector ${1 \over
2}(1 + \gamma_5 \!\!\not\!\! S)$ for $\ell^-$.

Regarding the expressions above, now we can define the single
lepton polarization. The definition of the polarized and
normalized differential decay rate is:
 \bea\label{diff}
\frac{d\Gamma^{\rho}(s,\vec{n})}{ds}=\frac{1}{2}\Bigg(\frac{d\Gamma^{\rho}}{ds}\Bigg)_0
[1+P^{\rho}_i \vec{e}.\vec{n}], \eea where a summation over
$i=L,\,T,\,N$ is implied. Polarized components $P^{\rho}_i$ in Eq.
(\ref{diff}) are as follows:
  \bea \label{e6312} P_{i}^{\rho} =
\frac{ {d\Gamma^{\rho}(\vec{n}=\vec{e}_i)}{d\hat{s} } -
 {d\Gamma^{\rho}(\vec{n}=-\vec{e}_i)}/{d\hat{s} }} {{d\Gamma^{\rho}(\vec{n}=\vec{e}_i)}/{d\hat{s} } +
 {d\Gamma^{\rho}(\vec{n}=-\vec{e}_i)}/{{d\hat{s}}}}. \eea
As a result the different components of the $P^{\rho}_i$ are
given as: \bea\label{pL}
P_L&=&\Bigg\{24~{\rm Re}(A^*B)(1-r_\rho-s)s(-1+v)+4m_B^2s\lambda_\rho v
~{\rm Re}(A^*D)\nnb
\\  &+&
\frac{1}{r_\rho}(3+v)\Bigg[2~{\rm Re}(B^*E)(1+r_\rho^2+2sr_\rho+s^2-2(r_\rho+s))\nnb
\\  &+&
m_B^2~{\rm Re}(C^*E)\{1-3(r_\rho+s)-(r_\rho-s)^2(r_\rho+s)+(3r_\rho^2+2sr_\rho+3s^2)\}\Bigg]\nnb
\\  &+&
\frac{1}{r_\rho}\{~{\rm Re}(B^*F)(1-r_\rho-s)+~{\rm Re}(C^*F)m_B^2\lambda_\rho\}\Bigg[(3+v)(1+r_\rho(r_\rho-s)-2r_\rho)\nnb
\\  &+&
(3-7v)s(r_\rho-s)-8sv\Bigg]\Bigg\}/\Delta_\rho
 \eea
 \bea\label{pT}
P_T=\frac{\pi t\sqrt{s\lambda_\rho}}{\Delta_\rho}\Bigg[-4~{\rm
Re}(A^*B)+\frac{1}{4sr_\rho}\{2(2(1-s-r_\rho)~{\rm
Re}(B^*E)m_B^2\lambda_\rho ~{\rm Re}(C^*E))\}\Bigg]
 \eea
 \bea\label{pN}
 P_N&=&\frac{\pi\sqrt{\lambda_\rho(s-4t^2)}}{\Delta_\rho}\Bigg[\frac{2(1+r_\rho-s)}{r_\rho}~{\rm Im}(E^*F)+2~{\rm
 Im}(A^*E+B^*D)\Bigg].
 \eea
These results for $P_L, P_T$ and $P_N$ agree with those given in
\cite{s.rai} when $\lambda_{t'}=0$. It also can be seen from the
explicit expression of $P_i^\rho$ involving various combination of
the Wilson coefficients that they are quite sensitive to the
fourth generation effects. Furthermore, $ P_N^\rho$ is
proportional to the imaginary parts of the product of the Wilson
coefficients. The existence of the new weak phase as a result of
 fourth generation contributes constructively to the magnitude
 of the $P_N^\rho$ .

\section{Numerical analysis}
In this section, we will study the dependence of the total
branching ratio, averaged CP asymmetry and lepton polarizations to
the mass of fourth quark ($m_{t'}$) and the product of quark
mixing matrix elements ($V_{t^\prime b}^\ast V_{t^\prime
d}=r_{sb}e^{i\phi_{db}}$). The main input parameters in the
calculations are the form factors. The definition of the form
factors are (see\cite{s.rai}):
 \bea
V(q^2)&=&\frac{V(0)}{1-q^2/5^2},\nnb
\\
A_1(q^2)&=&A_1(0)(1-0.023q^2),\nnb
\\
 A_2(q^2)&=&A_2(0)(1+0.034q^2),\nnb
\\
A_0(q^2)&=&\frac{A_3(0)}{1-q^2/4.8^2},\nnb
\\
A_3(q^2)&=&\frac{m_B+m_\rho}{2m_\rho}A_1(q^2)-\frac{m_B-m_\rho}{2m_\rho}A_2(q^2),\nnb
\\
 T_1(q^2)&=&\frac{T_1(0)}{1-q^2/5.3^2},\nnb
 \\
T_2(q^2)&=&T_2(0)(1-0.02q^2),\nnb
\\
T_3(q^2)&=&T_3(0)(1+0.005q^2).
 \eea with $V(0)=0.47, A_1(0)=0.37, A_2(0)=0.4, T_1(0)=0.19, T_2(0)=0.19, T_3(0)=-0.7$
We also use the SM parameters shown in table 1:
\begin{table}[h]
        \begin{center}
        \begin{tabular}{|l|l|}
        \hline
        \multicolumn{1}{|c|}{Parameter} & \multicolumn{1}{|c|}{Value}     \\
        \hline \hline
        $\alpha_{em}$                   & $1/129$ \\
        $m_{u}$                   & $2.3$ (MeV) \\
        $m_{d}$                   & $4.6$ (MeV) \\
        $m_{c}$                   & $1.25$ (GeV) \\
        $m_{b}$                   & $4.8$ (GeV) \\
        $m_{\mu}$                   & $0.106$ (GeV) \\
        $m_{\tau}$                  & $1.780$ (GeV) \\
        \hline
        \end{tabular}
        \end{center}
\caption{The values of the input parameters used in the numerical
          calculations}
\label{input}
\end{table}


 In order to perform quantitative analysis of the
total branching ratio, CP asymmetry and the lepton polarizations
the values of the new parameters ($m_{t'},\,r_{db},\,\phi_{db}$)
are needed. In the foregoing numerical analysis, we vary $m_{t'}$
in the range $175\le m_{t'} \le 600$GeV. The former is lower range
because of the fact that the fourth generation up quark is
expected to be heavier than the third generation ones ($m_t \leq
m_{t'}$)\cite{Sultansoy:2000dm}. The upper range comes from the
experimental bounds on the $\rho$ and $S$ parameters of SM,
furthermore, a mass greater than the 600GeV also contradicts with
partial wave unitarity~\cite{Sultansoy:2000dm}. As for mixing, we
use the result of Ref.~\cite{Hou:2005hd} where it was found that
$|V_{t'd}V_{t'b}|\sim 0.001$ with the phase about $10^\circ$,
which is consistent with the $sin2\phi_1$ of the CKM and the $B_d$
mixing parameter $\Delta m_{B_d}$~\cite{Hou:2005hd}.

 We can still move one more step further. From explicit expressions of the physical
observables, one can easily see that they depend on both $\ss$ and
the new parameters~($m_{t'},\,r_{db}$). One may eliminate the
dependence of the lepton polarization on one of the variables. We
eliminate the variable $\hat{s}$ by performing integration over
$\ss$ in the allowed kinematical region. The total branching
ratio, CP asymmetry  and the averaged lepton polarizations  are
defined as \bea {\cal B}_r&=&\ds \int_{4
m_\ell^2/m_{B}^2}^{(1-\sqrt{\hat{r}_{\rho}})^2}
 \frac{d{\cal B}}{d\hat{s}} d\hat{s},
\nnb\\\lla P_{i}(A_{CP}) \rra &=& \frac{\ds \int_{4
m_\ell^2/m_{B}^2}^{(1-\sqrt{\hat{r}_{\rho}})^2} P_i(A_{CP})
\frac{d{\cal B}}{d\hat{s}} d\hat{s}} {{\cal{B}}_r}~. \eea

Figs. (1)--(6) depict the dependence of the total branching ratio,
unpolarized CP asymmetry and averaged lepton polarization for
various  $r_{db}$ in terms of  $m_{t'}$. We should note here that
as the dependency for various $\phi_{db}\sim\{0^\circ-30^\circ\}$
is too weak we just show the result only for $\phi_{db}=15^\circ$.
Also, we do not present  the deviation of some  observables for
which the corresponding three-generation standard model values are
less than $1\%$, i.e. $ \langle P_N^\rho\rangle$ for muon and tau
leptons and $\langle A_{CP}^\rho\rangle, \,\,\langle
P_N^\rho\rangle$ for tau lepton.

From the figures it can be concluded that:

\begin{itemize}

\item ${\cal B}_r$  strongly depends on the mass of the fourth
quark~($m_{t'}$)
 and  quark mixing matrix product~($r_{db}$) for both $\mu$
 and $\tau$ channels. Furthermore, for both channels, ${\cal B}_r$ is enhanced sizably in terms of  both $m_{t'}$ and $r_{db}$.

\item $P_L^\rho$ and $ A_{CP}^\rho$ are independent from the
lepton mass~(See Eq. (\ref{pL}) and (\ref{acp})). For this reason,
considering
 a fixed value of $\ss$, they are the same for $e,\,\mu\,\mbox{and }\tau$
channels. The situation  for the $\lla P_L^\rho\rra$ and $\lla
A_{CP}^\rho\rra$ is different; those values for the $\tau$ channel
are less than for the $\mu$ channel. This is because of the fact
that the phase integral space is suppressed by increasing the
lepton mass~($m_\ell$). The SM3 value of $\lla P_L^\rho\rra$ and
$\lla A_{CP}^\rho\rra$ almost vanishes for the $\tau$ channel. The
SM4 suppresses those values even more. On the other hand, $\lla
P_L^\rho\rra$ and $\lla A_{CP}^\rho\rra$ for the  $\mu$ channel
 strongly depend on the SM4 parameters. Magnitude of both is a
decreasing function of the $r_{db}$ and $m_{t'}$.

\item  $\lla P_T^\rho\rra$ strongly  depends on the fourth quark
mass($m_{t'}$) and quark mixing matrix product($r_{sb}$) for both
the $\mu$ and $\tau$ channels. Its magnitude is a decreasing
function of both $m_{t'}$ and $r_{sb}$.
 The measurement of the  magnitude and sign of this observable can be used as
a good tool to search for fourth generation effects.
\end{itemize}

To sum up, we presented the  systematic analysis of the $B \rar
\rho \ell^- \ell^+$ decay by using the SM with fourth generation.
 The sensitivity of the total branching ratio, CP
asymmetry and lepton polarization on the new parameters, which
come out of fourth generation, were studied. We found out that
above-mentioned physical observables  depict a strong dependence
on the fourth quark~($m_{t'}$) and on the product of quark mixing
matrix elements~($V_{t^\prime b}^\ast V_{t^\prime
d}=r_{db}e^{i\phi_{db}}$). We found that while the branching ratio
and $\lla P_N^\rho\rra$ are enhanced, CP asymmetry, $\lla
P_L^\rho\rra$ and $\lla P_T^\rho\rra$ are suppressed by fourth
generation effects.
 The measurement of the magnitude and sign of these
readily measurable observables, in particular for the $\mu$  case,
can serve as a good tool to search for physics beyond the SM. In
particular, the results can be used for an indirect search to look
for the fourth generation of quarks.
%

\newpage

\section*{Figure captions}

{\bf Fig. (1)} The dependence of the branching ratio of $B\rar
\rho \mu^+\mu^-$ on $m_{t'}$ for
 $r_{db}=0.001,\, 0.002,\,0.003$\\ \\
{\bf Fig. (2)} The same as in Fig. (1), but for the $\tau$ lepton\\ \\
{\bf Fig. (3)}  The dependence of the $\lla A_{CP}\rra$ of $B\rar
\rho \mu^+\mu^-$  on $m_{t'}$ for
 $r_{db}=0.001,\, 0.002,\,0.003$\\ \\
{\bf Fig. (4)} The dependence of the $\lla P_L\rra$ for $\mu$
lepton on $m_{t'}$ for
 $r_{db}=0.001,\, 0.002,\,0.003$\\ \\
 {\bf Fig. (5)} The same as in Fig. (5), but for the $\tau$ lepton\\
 \\
 {\bf Fig. (6)} The dependence of the $\lla P_T\rra$ for $ \mu $ lepton,
on $m_{t'}$ for
 $r_{db}=0.001,\, 0.002,\,0.003$\\ \\
  {\bf Fig. (7)} The same as in Fig. (6), but for the $\tau$ lepton\\
 \\
\newpage
\begin{figure}
\vskip 1.5 cm
    \includegraphics{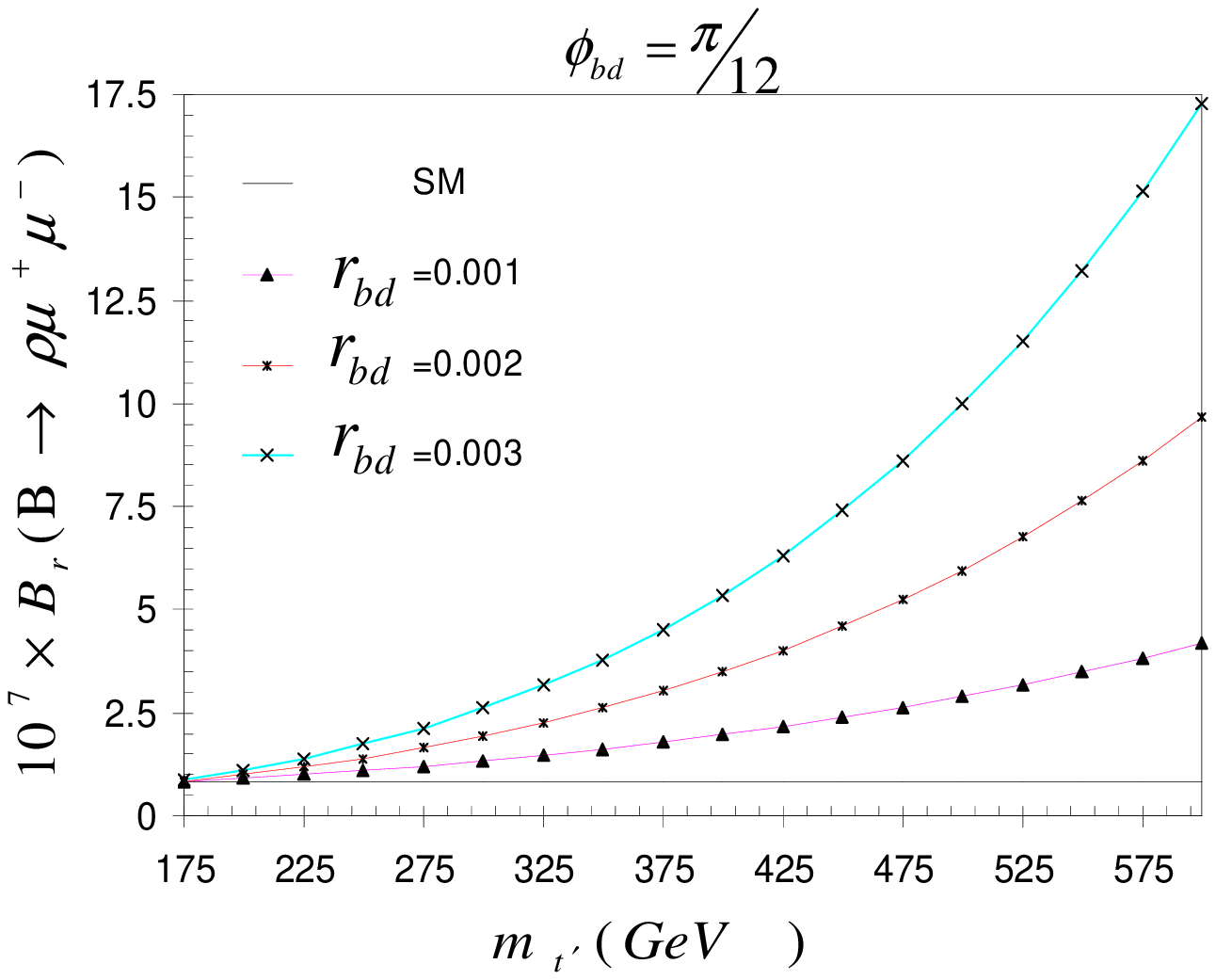}
\vskip 6.5cm \caption{}
\end{figure}
\begin{figure}
\vskip 1.5cm
    \includegraphics{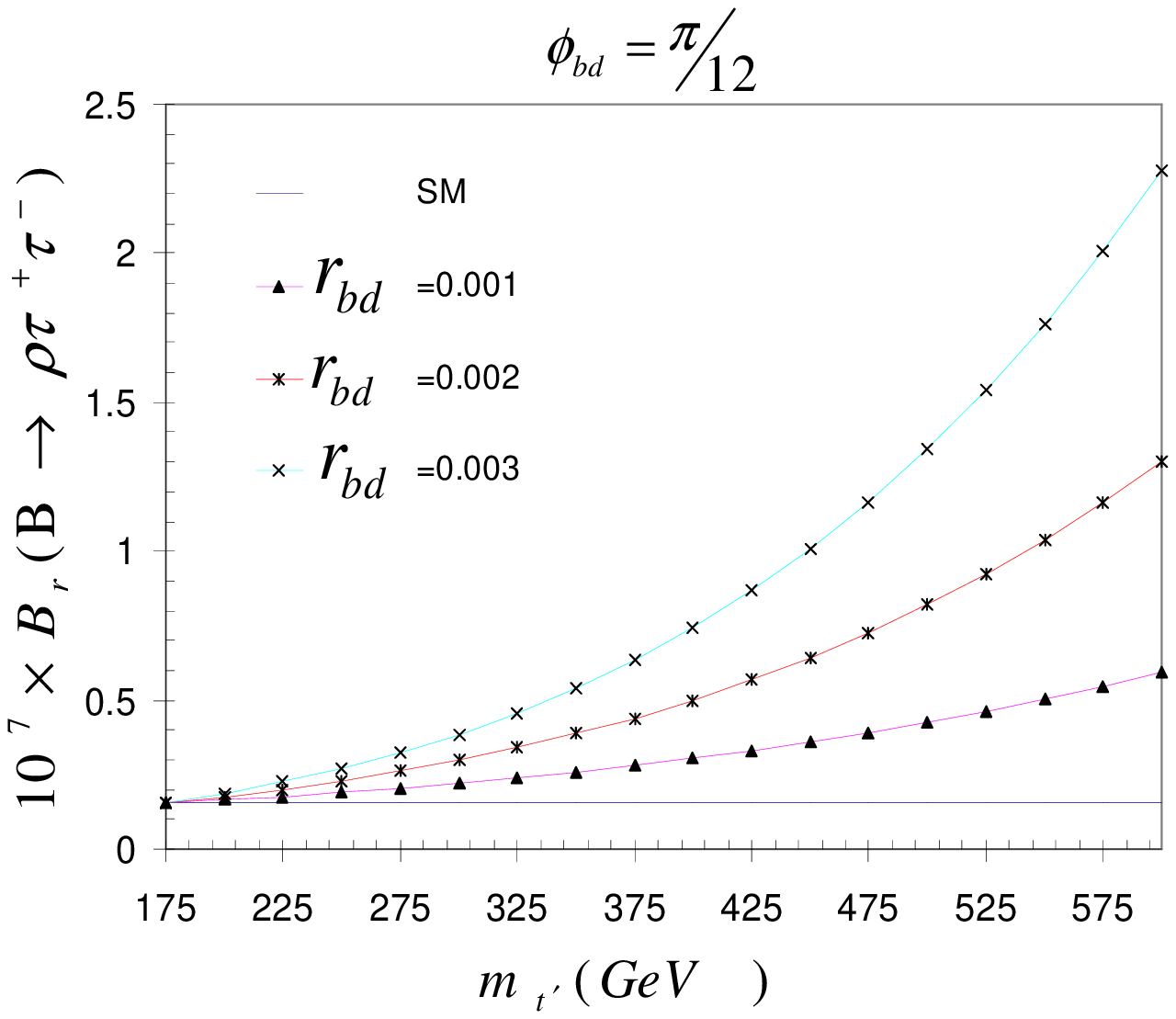}
\vskip 6.5cm \caption{}
\end{figure}
\begin{figure}
\vskip 1.5 cm
    \includegraphics{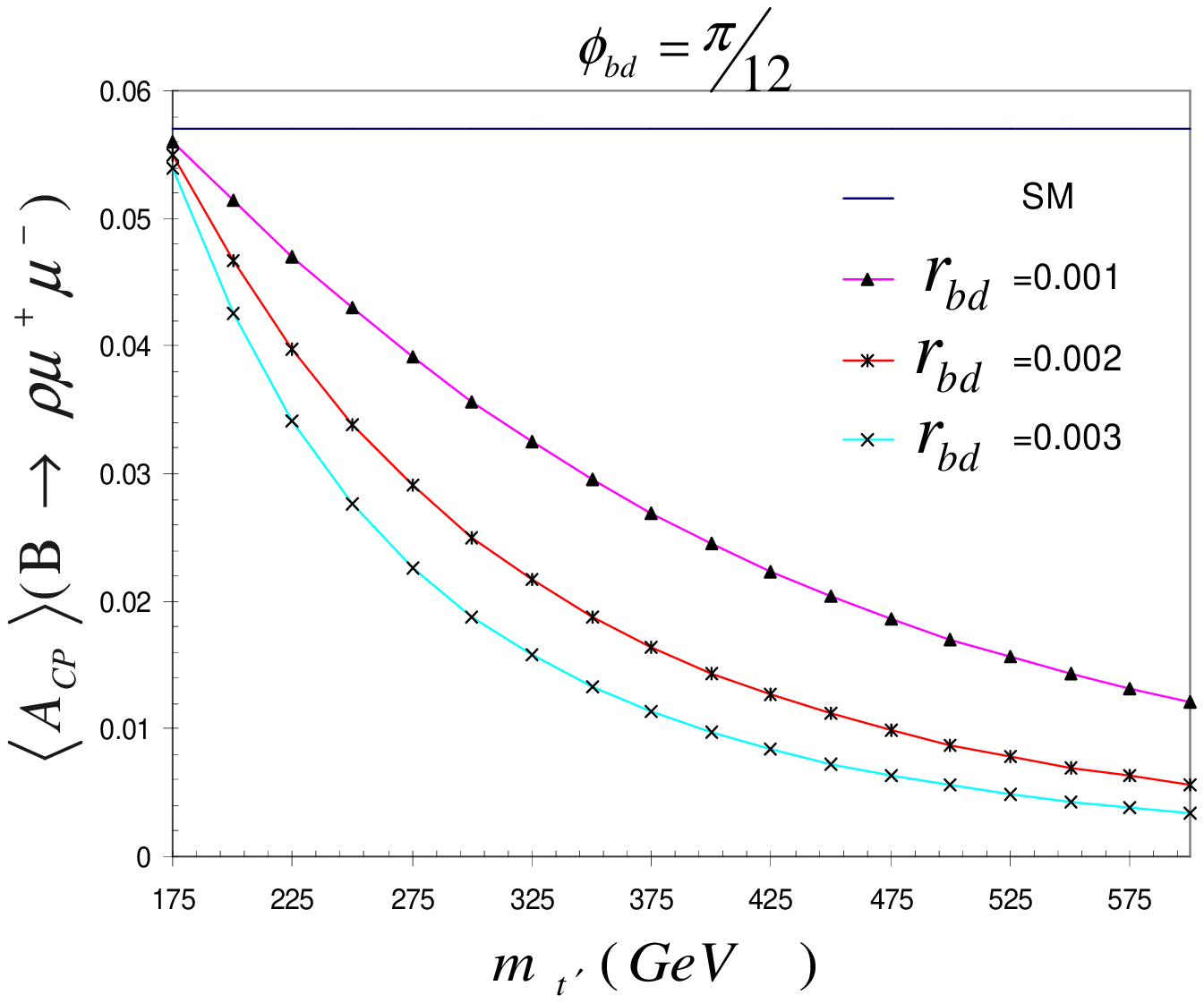}
\vskip 7.8cm \caption{}
\end{figure}
\begin{figure}
\vskip 1.5 cm
    \includegraphics{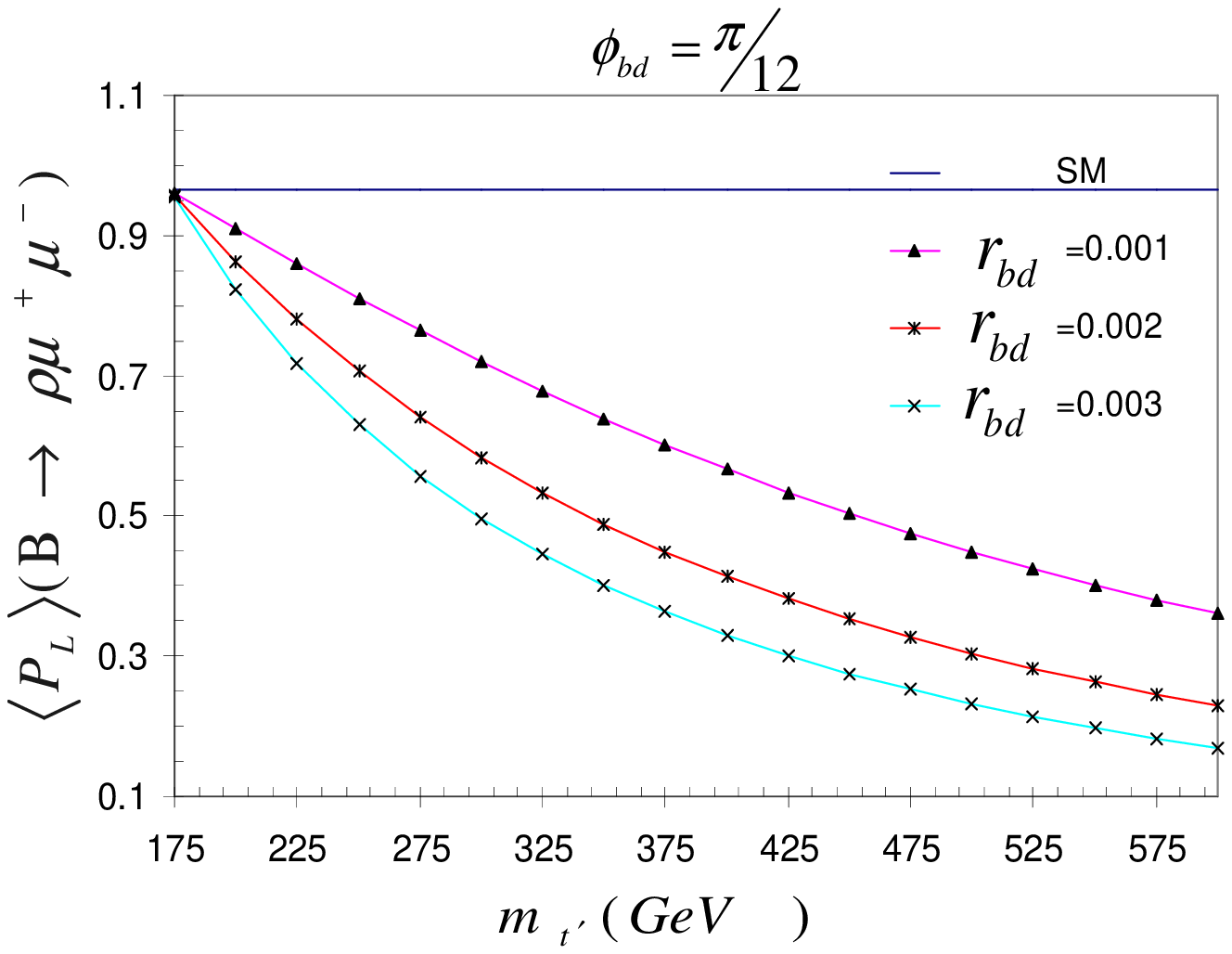}
\vskip 7cm \caption{}
\end{figure}
\begin{figure}
\vskip 1.5 cm
    \includegraphics{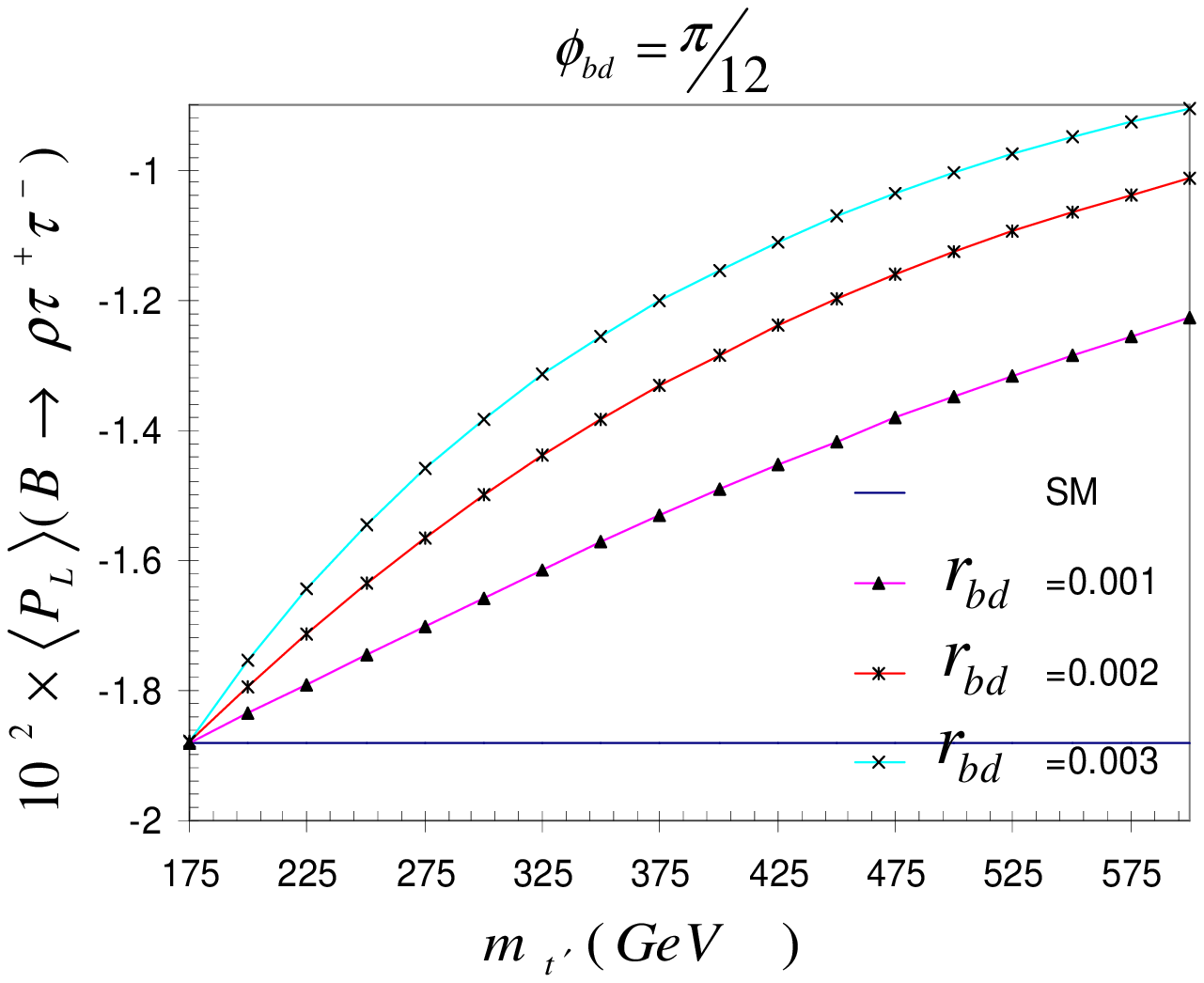}
\vskip 7.8cm \caption{}
\end{figure}
\begin{figure}
\vskip 1.5cm
    \includegraphics{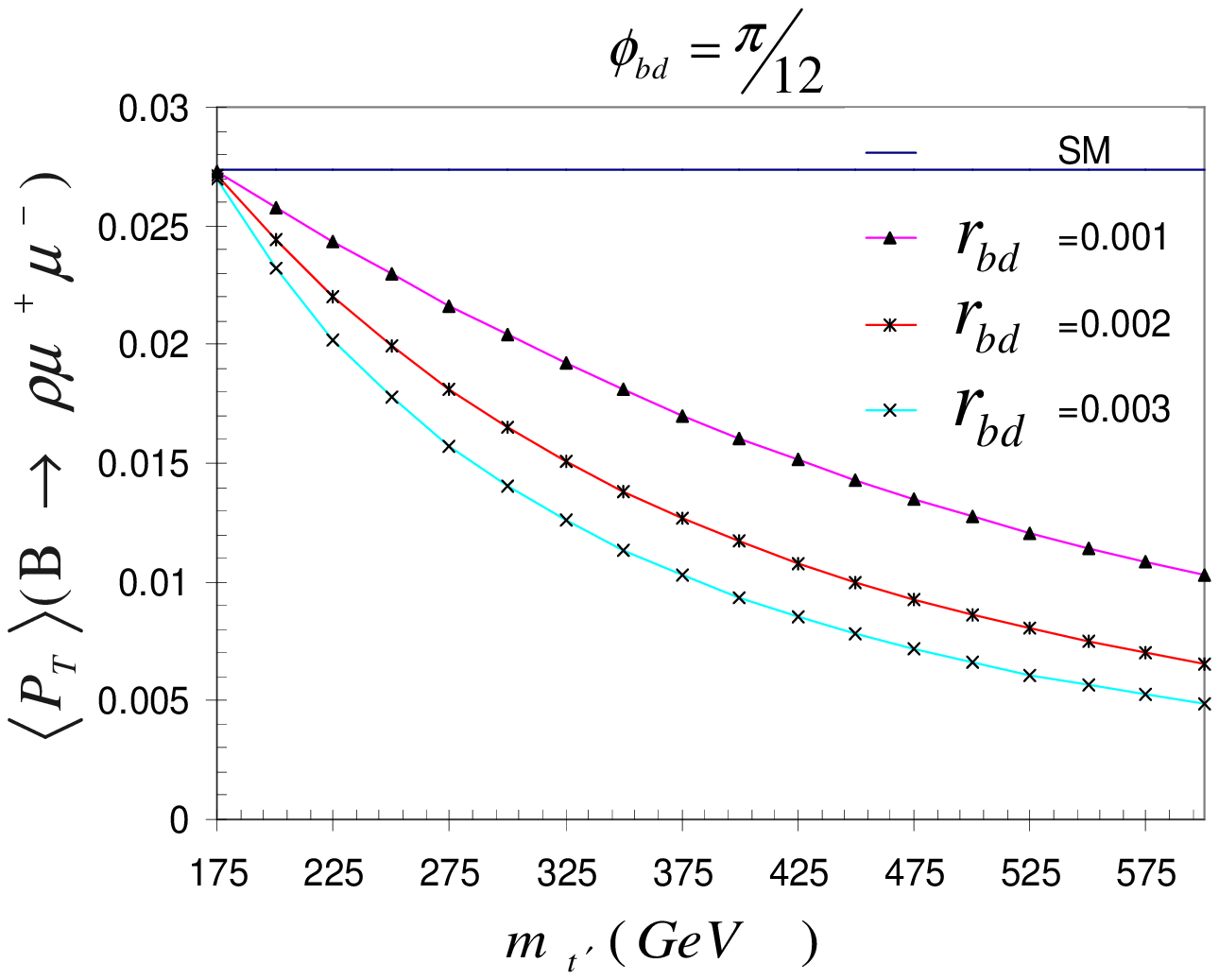}
\vskip 6.8cm \caption{}
\end{figure}
\begin{figure}
\vskip 1.5 cm
    \includegraphics{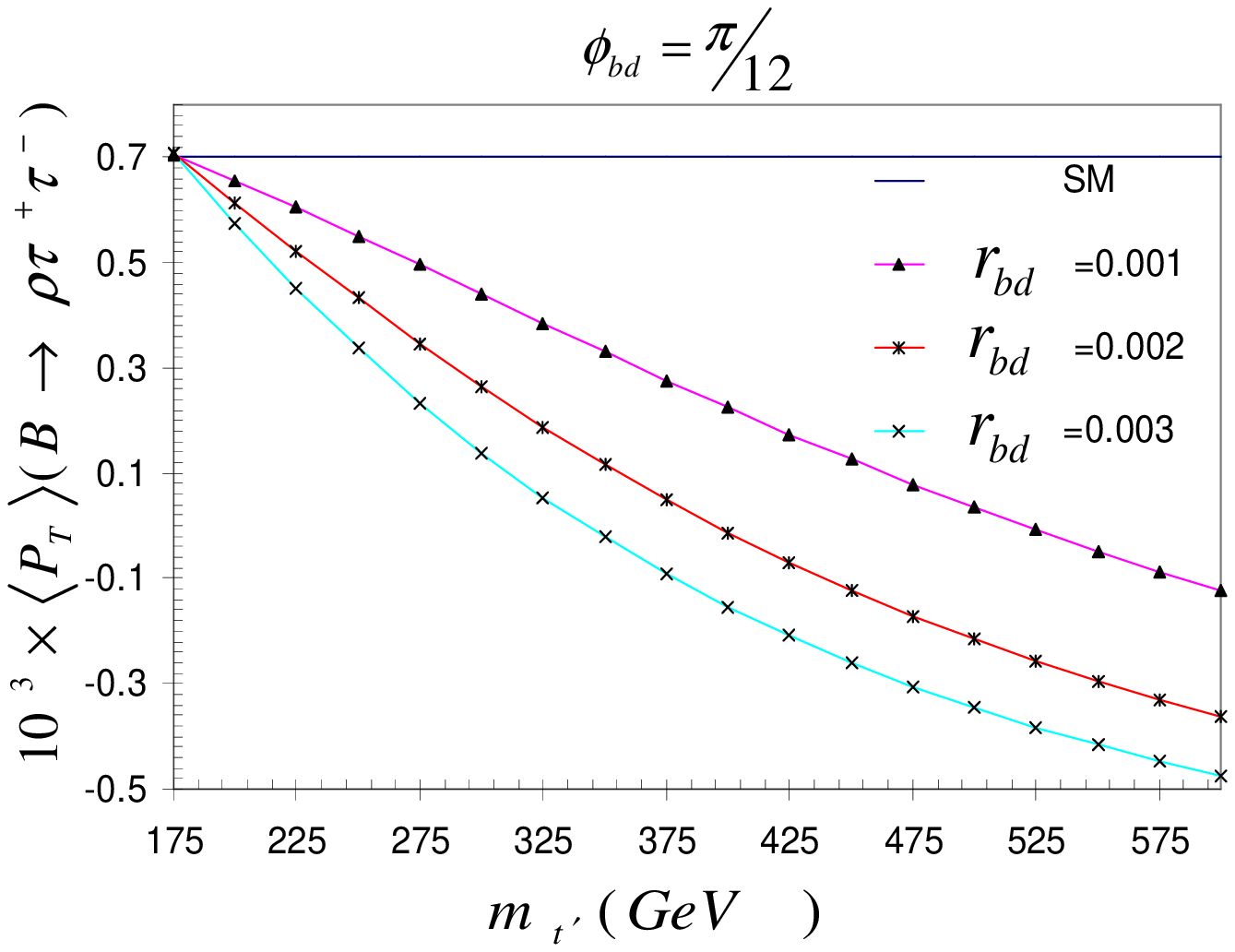}
\vskip 6.5cm \caption{}
\end{figure}

\end{document}